\documentclass[12pt]{article}
\input psfig.tex
\begin{document}

\begin{center}{\large\bf Electromagnetic Polarization Effects 
due to\\ 
Axion Photon Mixing}
\end{center}
\bigskip
\centerline{\bf Pankaj Jain, Sukanta Panda
and S. Sarala}
\bigskip
\begin{center}
Physics Department, I.I.T. Kanpur, India 208016\\
\end{center}

\bigskip
\noindent{\bf Abstract:} We investigate the effect of axions on the 
polarization of electromagnetic waves as they propagate through 
astronomical distances. We analyze the change in the
dispersion of the electromagnetic wave due to its mixing with axions.
We find that this leads to a shift in polarization and turns out 
to be the dominant effect for a wide range of frequencies.
We analyze whether this effect or the decay of
photons into axions can explain the large scale anisotropies which
have been observed in the polarizations of quasars and radio galaxies.
We also comment on the possibility that the axion-photon mixing can explain
the dimming of distant supernovae.

\section{Introduction}
The mixing of axions with photons and its observable consequences
have been analyzed by many authors [1-12]. 
In the present paper
we investigate the changes in the polarization of electromagnetic waves that
arise due to its mixing with axions. We are particularly interested
in determining if this mixing can explain the polarization anisotropies
that have been claimed in Ref. \cite{Birch,JR,Hutsemekers}. 

In Ref. \cite{Birch,JR} the authors claimed that the observed polarizations
from distant radio galaxies and quasars are not isotropically distributed
on the dome of the sky. The observable of interest in that study 
was the angle $\beta=\chi-\psi$,
where $\psi$ is the orientation angle of
the axis of the radio galaxy and $\chi$ the observed polarization angle
after the effect of Faraday rotation is taken out of the data.
The authors claimed a dipole anisotropy such that the angle 
$\beta$ is given by
\begin{equation}
\beta = \vec\lambda\cdot \hat r
\end{equation}
where $\hat r$ is a unit vector in the direction of the source. The 
$\vec\lambda$ represent the three parameters of this fit. The magnitude
of this vector $|\vec\lambda|$ is found to be approximately 0.5 and
its direction 
\begin{equation}
\hat \lambda = [(0\ h, 9\ m)\pm (1\ h,0\ m),-1^o\pm 15^o]\ .
\label{axis}
\end{equation} 
The effect is independent of redshift and was first claimed in Ref. 
\cite{Birch} and later verified by a more reliable statistical
procedure \cite{KY} and by compiling a larger data set \cite{JR}. 
This anisotropy may be a signal of 
some local effect arising due to the milky way or the local supercluster.
However so far it is not known what physical phenomenon could lead 
to the observed rotation in polarizations. Within the standard model
of elementary particles
it is difficult to conceive of a physical mechanism which can lead
to this effect. The axion field which arises
in many extensions of the standard model of particle physics 
may provide one possible explanation. However so far it is not known whether
such a field can consistently explain this effect. 
This is one of the questions that we study in the present paper.

Another interesting polarization effect in the electromagnetic waves from
distant quasars has been claimed in Ref. \cite{Hutsemekers,HL}.
It was found that optical polarization are aligned  on very large scales.
A very striking alignment was found in the region, called A1 in
\cite{HL}, delimited in Right Ascension by $11^{\rm h}15^{\rm m}
\le {\rm RA} \le 14^{\rm h}29^{\rm m}$ and in redshift by $1.0\le z\le 2.3$.
The polarizations from quasars in any particular
spatial region have a tendency to align with one another. The effect
was only seen in patches without any evidence of large scale anisotropy.
We point out that the center of the A1 region (see Fig. 1 of 
Ref. \cite{HL}) is exactly opposite to
the axis, Eq. \ref{axis}, of the anisotropy found in Ref. \cite{JR}. 
This might indicate a common origin of these two effects. 

Finally we also examine the recent claim \cite{Csaki} 
that dimming of distant supernovae \cite{Perlmutter,Riess}
can be explained in terms of axion photon mixing.  

\section{Axion-Photon Mixing}
The interaction lagrangian of the axions with electromagnetic field can be 
written as \cite{Krauss}
\begin{equation}
{\cal L}_{\rm int} = {\alpha N\over 12\pi f_a}\phi F_{\mu\nu}\tilde F^{\mu\nu}
\end{equation}
where $\phi$ is the axion field, $F_{\mu\nu}$ is the electromagnetic
field tensor, $f_a$ is the scale of PQ \cite{PQ} symmetry breaking and $N$ is
the number of light quark flavors. 
The current limits on $f_a$ 
 are given by
\begin{equation}
f_a > 10^{10} \ {\rm GeV}
\end{equation} 
The axion mass
is related to the coupling $1/f_a$ by
\begin{equation}
m_a^2 =N^2{f_\pi^2 m_\pi^2\over f_a^2} {m_u m_d\over (m_u+m_d)^2} 
\label{Axion_Mass}
\end{equation}
where $f_\pi$ is the pion decay constant, $m_\pi$ the pion mass, 
and $m_u$ and $m_d$ the masses of up and down quarks respectively.
It is also interesting to consider other pseudoscalar particles which
arise in certain extensions of standard model, whose mass
is not related to $f_a$ by Eq. \ref{Axion_Mass}.
In most of our discussion below we will take the mass of the pseudoscalar
as a free parameter and not given by Eq. \ref{Axion_Mass}. 

Axions can mix with photons from distant galaxies and lead to rotation
of polarization. The basic picture is that the photons emitted
by the galaxy can decay into axions as they propagate through the 
background magnetic
field. Since only the photons polarized parallel to the
transverse component of the background magnetic
field ($\vec B_t$) decay, this effect can lead to a 
change in polarization of the 
electromagnetic wave. Another effect that can also contribute is the
mixing of photon with an off-shell axion.
This process changes the dispersion relation for the photon
polarized parallel to  $\vec B_t$ 
and hence can lead to  
changes in polarization.  
Alternatively the distant galaxies may be emitting axions along with
photons. These axions can decay in the presence of magnetic fields into
photons which are polarized parallel to  $\vec B_t$. 
We examine all of these possibilities. 

The basic equations can be written as \cite{RS88,CG94},
\begin{equation}
(\omega^2 + \partial_z^2 - \omega_p^2)A_\parallel + gB_t\omega\phi = 0
\label{waveEq1}
\end{equation}
\begin{equation}
(\omega^2 + \partial_z^2 - m_a^2)\phi + gB_t\omega A_\parallel = 0
\label{waveEq2}
\end{equation}
where 
$\omega_p$ is the plasma frequency,
$A_\parallel$ is the component of the vector potential parallel $\vec B_t$, 
$B_t=|\vec B_t|$
and the coupling $g = {\alpha N\over 3\pi f_a}$.
By making the ansatz,
\begin{equation}
A^\prime_\parallel = \exp\left(i\omega z-i\int_0^z {\omega_p^2(z^\prime)\over
2\omega}dz^\prime\right) A_\parallel
\end{equation}
\begin{equation}
\phi^\prime = \exp\left(i\omega z-i {m_a^2z\over
2\omega}\right)\phi
\end{equation}
and assuming that the frequency $\omega$ is much larger than the mass 
eigenvalues, 
the wave equations can be written as,
\begin{equation}
-i\partial_z A^\prime_\parallel + {gB_t\over 2}
\exp(-i\xi(z)/2\omega)\phi^\prime 
= 0
\label{waveEq3}
\end{equation}
\begin{equation}
-i\partial_z \phi^\prime + {gB_t\over 2}\exp(i\xi(z)/2\omega)A^\prime_\parallel 
= 0
\label{waveEq4}
\end{equation}
where
\begin{equation}
\xi(z) = \int_0^z dz^\prime (\omega_p^2(z^\prime)- m_a^2) 
\end{equation} 
In arriving at equations \ref{waveEq3} and \ref{waveEq4} we have ignored
second derivatives of $A^\prime_\parallel$ and $\phi^\prime$ since these
are slowly varying fields.

We first replace the $\omega_p^2(z)$ 
with its mean value ${\overline\omega}_p^2$.
In this case, as shown in Ref. \cite{CG94}, the probability for producing
the pseudoscalar particles $|\phi(L)|^2$ at distance $L$, assuming that 
$\phi(0)=0$ is equal to   
\begin{equation}
P_{\gamma\rightarrow \phi} = |\phi(L)|^2 = |\phi^\prime(L)|^2
\approx (gB_tl)^2
\sin^2\left[{L\over 2l}\right]
\label{ProbA}
\end{equation}
where 
\begin{equation}
l = {2\omega\over \overline\omega_p^2 - m_a^2}
\ .
\end{equation}
This 
gives a very small value assuming the current limit for the value
of $g$ and the galactic values 
for $B_t$ and $\overline\omega_p$, independent of the choice of
$m_a$. The special case where $\overline\omega_p^2$ is accidently 
equal to  $m_a^2$
is not considered in this paper. 
The effect is, however, much larger if the variations in the plasma
frequency and/or background magnetic field are taken into account
\cite{CG94}. The authors in Ref. \cite{CG94} 
assume a Kolmogorov power spectrum,
$P_{3N} = C_N^2k^{-11/3}$ for the 
electron density fluctuations, where $C_N\approx 3\times 10^{-4}$ 
m$^{-20/3}$ for the interstellar medium. 
The authors then show that for the electromagnetic
wave propagating through interstellar medium the probability to
produce axions, $P_{\gamma\rightarrow \phi}$, is given by,
\begin{eqnarray}
P_{\gamma\rightarrow \phi} &\approx & 2.7\times 10^{-5} \left[{g\over 10^{-10}\ 
{\rm GeV}^{-1}}\right]^2\left({\overline B_t\over \mu {\rm G}}\right)^2
\left({0.03
{\rm cm}^{-3}\over \overline n_e}\right)^{11/3}\nonumber\\
& \times &
\left({C_N^2\over 3\times 10^{-4}\ {\rm m^{-20/3}}}\right)\left({L\over
{\rm kpc}}\right)\left({\nu\over 10^6\ {\rm GHz}}\right)^{5/3}
\left|1-{m_a\over\overline\omega^2_p}\right|^{-11/3}
\label{Prob}
\end{eqnarray}
In obtaining this result the magnetic field has been assumed to be
constant. We have also assumed that the vector potential is approximately
constant and kept only the leading order term
in the expansion in powers of the fluctuations. Hence this result is valid
only if the conversion probability is small. 
The conversion probability given by this equation is much
larger than that implied by Eq. \ref{ProbA}. 
For propagation over galactic distances
this probability is
still very small compared to unity unless the frequency of the electromagnetic
wave is much larger than the optical frequencies.
However for supercluster magnetic fields the probability
may be equal to unity even at optical frequencies. For example, in the
Virgo supercluster of galaxies the magnetic field strength is found to
be about 1 $\mu G$ over a very large length scale of 10 Mpc with
plasma density of the order of $10^{-6}$ cm$^{-3}$ \cite{Vallee}. In order to
compute the correlation coefficient $C_N$ for the supercluster plasma
density we assume, by dimensional
analysis, that it scales as 
$n_e^{20/9}$, since $n_e$ is the only dimensionful parameter known. 
This gives $C_N^2\approx 3.4\times 10^{-14}\ {\rm m}^{-20/3}$.
This estimate of $C_N$ is not reliable and must be improved in future
by direct observations. Using this rough estimate 
the probability is found to be approximately $8\times 10^5$ for $g=10^{-10}$
GeV$^{-1}$. This is large enough that it can also significantly
affect the CMBR in the direction of the Virgo supercluster. By demanding
that $P_{\gamma\rightarrow\phi} < 10^{-5}$ for the microwave photons
$\nu = 50 $ GHz we find that $g < 0.9\times 10^{-11}$ GeV$^{-1}$. This
constraint is comparable to the constraint on coupling $g$ obtained
from red giants and SN1987A \cite{Rosenberg,Raffelt,Brockway} 
but depends on our assumed extrapolation of $C_N$ and hence
is not entirely reliable. The constraint imposed by CMBR also implies that 
\begin{equation}
P_{\gamma\rightarrow\phi} < 6.5\times 10^3 \left({\nu\over 10^6\ {\rm GHz}}\right)^{5/3} 
\label{CMBR}
\end{equation}
The conversion probability is ofcourse always less than one. The large 
conversion probability is obtained by using Eq. \ref{Prob} in the 
regime where
it is not valid. As mentioned above, Eq. \ref{Prob} is applicable
only when the conversion probability is small. 
In the present case the result, Eq. \ref{CMBR}, 
has to interpreted in the sense that 
for a considerable range of parameters 
$P_{\gamma\rightarrow \phi}$ is of order unity 
for $\nu\ge 10^6$ GHz.

We next investigate how the polarization of photons will change due 
to change in the dispersion of photon because of  
their mixing with axions. 
We integrate Eq. \ref{waveEq4} by parts, taking   
$B_t$ as constant, to obtain 
\begin{equation}
\phi^\prime(z) - \phi^\prime(0) = 
-{gB_t\over 2}l \left[e^{iz/l}A^\prime_\parallel(z) 
- A^\prime_\parallel(0) - \int_0^z dz' \partial_z A^\prime_\parallel\exp(iz^\prime/l)\right]  
\ .
\label{soln_phi}
\end{equation}
It is clear that the third term inside the brackets is higher order
in $g$ and can be dropped. 
Substituting this expression for $\phi^\prime$ into Eq. \ref{waveEq3}
we find that
\begin{equation}
\partial_z A^\prime_\parallel(z) = -i{gB\over 2}\phi^\prime(0)\exp(-iz/l) 
+ i{g^2B_t^2\over 4}l
\left[A^\prime_\parallel(z) -\exp(-iz/l) A^\prime_\parallel(0)\right]\  .
\end{equation}
The solution to this equation can be written as,
\begin{eqnarray}
A^\prime_\parallel(L) &=& A^\prime_\parallel(0)\exp(i\Gamma(L)) 
- \left({l\over 1+l\Gamma(L)/L}\right)\left[{gB_t\over 2}\right] 
\left(\phi^\prime(0)
+{gB_t\over 2} lA^\prime_\parallel(0)\right)\nonumber \\
&\times &
\left[2\sin^2\left(
{L\over 2l} + {\Gamma(L)\over 2}\right) + i \sin\left({L\over l}
+ \Gamma(L)\right)\right]\exp(i\Gamma(L))
\label{Omega}
\end{eqnarray}
where,
\begin{equation}
\Gamma(L) = \left({gB_t\over 2}\right)^2lL
\end{equation}
For large $L$, of the order of galactic or super-galactic scales, 
the dominant contribution comes from the first term on
the right hand side of Eq. \ref{Omega}. 
This term leads to an additional phase $\Gamma$ 
for the parallel component of the 
vector potential in 
comparison to the perpendicular component. 
The total photon flux fluctuates due to its conversion into axions. 

In the optical frequencies, assuming the
galactic values of the magnetic field $B_t\approx 1 \mu G$ and the plasma 
density $\overline n_e\approx 0.03$ cm$^{-3}$, 
we find that $\Gamma(L)$ is of order of $10^{-5}$ for $L\approx 1$ kpc
and $g\approx 10^{-10}$ GeV$^{-1}$ which is comparable to
the probability of conversion of photon into an axion as given by Eq.
\ref{Prob}. However for smaller frequencies this gives a much larger result
in comparison to Eq. \ref{Prob}.
If we assume the Virgo supercluster values of the magnetic field 
$B_t\approx  1 \mu G$
and plasma density $\overline n_e\approx 10^{-6}$ cm$^{-3}$ \cite{Vallee}
we find that
for $m_a<< \overline \omega_p$,
\begin{equation}
l \approx  {\nu\over 10^6\  {\rm GHz}}\ 0.04\ {\rm Mpc}
\label{plasmal}
\end{equation}
and the phase
$\Gamma$ can be written as
\begin{eqnarray}
\Gamma(L) &=& 2.4\times 10^3 
\ \left[{g\over 10^{-10}\ {\rm GeV^{-1}}}\right]^2\left({\overline B_t\over \mu {\rm G}}\right)^2
\left({10^{-6}
{\rm cm}^{-3}\over \overline n_e}\right)\nonumber\\
&\times &\left({L\over 10\ {\rm Mpc}}
\right)\left[{\nu\over 10^6\ {\rm GHz}}\right]\ .
\end{eqnarray} 
This phase can produce 
observable changes in the polarization of the electromagnetic wave
as we discuss later.   

We next compute the contribution to the phase that arises due to 
fluctuations in the plasma density. We assume, for simplicity, that
the background magnetic field is constant. Then solving for 
$\phi^\prime$ in Eq. \ref{waveEq4} and substituting into Eq. 
\ref{waveEq3} we
find,
\begin{equation}
\partial_z A_\parallel^\prime = -\left({gB_t\over 2}\right)^2
\exp\left(-i{\xi\over 2\omega}\right)\int_0^z dz^\prime 
\exp\left(i{\xi\over 2\omega}\right) A_\parallel^\prime
\end{equation}
Here we have set $\phi^\prime(0)=0$. If $\phi^\prime(0)\ne 0$ then
we will find another term on the right hand side which can be evaluated
easily to leading order in the fluctuations. We discuss this case later.  
We next write $A_\parallel^\prime = e^{i\Omega} \approx 1+i\Omega$,
where $\Omega$ is in general complex. We set $A_\parallel^\prime(0) = 1$
and hence $\Omega(0) = 0$. We assume that $\Omega(L)$ 
is small and keep only the leading order terms. In order to compute the
higher order terms we will require the higher order 
density correlations functions
for the turbulent interstellar (or intergalactic) medium. These
correlation functions are, however, unknown and hence it is not
possible to go beyond the leading order term. In any case
even the leading order
calculation is very useful since it can reliably provide
us with the parameter ranges where the axion photon mixing 
effect may be significant.
We find that
$\Omega(L)$ can be written as
\begin{equation}
\Omega(L) = i \left({gB_t\over 2}\right)^2 \int_0^L dz\exp\left(-i{\xi\over 
2\omega}\right)\int_0^z dz^\prime
\exp\left(i{\xi\over 2\omega}\right)
\label{EqOmega}
\end{equation}
We can now expand $\xi/2\omega$ in terms of the fluctuations in the 
plasma density to get
\begin{equation}
{\xi\over 2\omega} = {z\over l} + {1\over 2\omega} {\overline\omega_p^2\over n_e}
\int_0^z dz^\prime
\delta n_e
\end{equation}
where we have used $\omega_p^2 \propto n_e$, $\omega_p^2 = \overline
\omega_p^2 + \delta \omega_p^2$ and $n_e = \overline n_e + \delta n_e$. 
Substituting this into Eq. \ref{EqOmega} and expanding the right hand side
to second order in the fluctuations $\delta n_e$ we can write the
expectation value $<\Omega> = \Omega_0 + \delta\Omega$ with 
\begin{equation}
\delta\Omega(L) = \left({gB_t\over 2}\right)^2 {2il^2\over 8\omega^2}
\left({\overline\omega_p^2\over \overline n_e}\right)^2 e^{-iL/l}I_1 + 
\left({gB_t\over 2}\right)^2{il^2\over 4\omega^2}
\left({\overline\omega_p^2\over \overline n_e}\right)^2  (I_2-I_3)
\label{ExpOmega}
\end{equation}
where,
\begin{eqnarray}
I_1 &=& \left<\int_0^Ldz e^{iz/l} \delta n_e\int_0^z
dz_1 \delta n_e \right> \nonumber\\
I_2 &=& \left<\int_0^Ldz e^{-iz/l} \delta n_e\int_0^z
dz_1 e^{iz_1/l} \delta n_e \right> \nonumber\\
I_3 &=& \left<\int_0^Ldz e^{-iz/l} \delta n_e\int_0^z
dz_1 \delta n_e \right> 
\end{eqnarray}
and $\Omega_0$ is the contribution obtained by keeping only the mean
electron density and can be extracted from Eq. \ref{Omega}.

These integrals can be evaluated by expressing them in terms
of the Fourier transform of $\delta n_e=\int d^3{\bf k} \tilde n_e({\bf k})
e^{i{\bf k}\cdot {\bf x}}$ and using the Kolmogorov power spectrum 
$\left<\tilde n_e({\bf k})\tilde n_e^*({\bf k^\prime})\right> = P_{3N}(k)
\delta^3({\bf k} - {\bf k^\prime})$, where $k=|{\bf k}|$ and 
\begin{equation}
P_{3N}(k) = {C_N^2\over (k^2 + L_0^{-2})^{11/6}} \exp\left(-k^2l_0^2/2\right).
\label{Kolmogorov}
\end{equation}
Here $C_N^2$ is the correlation coefficient, $l_0$ is the inner scale
and $L_0$ the outer scale \cite{Armstrong}. 
In our calculation we set $l_0=0$. We
have explicitly kept the dependence on the outer scale $L_0$ since 
we find that some of the integrals are infrared divergent which are
regulated by this scale. 
We point out that since $\delta n_e$ is real $\tilde n_e^*({\bf k})
= \tilde n_e(-{\bf k})$. 
We find that $\delta\Omega$ is given by
\begin{eqnarray}
\delta\Omega(L) &\approx& i{6\over 5}\pi^2C_N^2{l^2\over 4 \omega^2}
\left({gB_t\overline\omega^2_p\over 2n_e} \right)^2 Ll^{5/3}
+ {6\over 5}\pi C_N^2 {l^2\over 4 \omega^2} 
\left({gB_t\overline\omega^2_p\over 2n_e} \right)^2\nonumber\\
&\times &\int_{-\infty}^\infty dk_z {1\over \left(k_z^2 + 1/L_0^2 \right)^{5/6}}
{1\over {1\over l} - k_z} \left[L - {\sin(L/l-k_zL)\over {1\over l} - k_z} 
\right]
\label{DeltaOmega}
\end{eqnarray}
In obtaining this result we have kept only the dominant terms which 
scale like $L$ in the limit of large $L>> l, L>> L_0$. The imaginary
part $\Omega_I$ of $<\Omega>= \Omega_R + i\Omega_I$, gives the 
decay rate of the photons with polarization vectors parallel 
to the transverse component of background magnetic field. It is
related to the production rate of axions by $|\phi^\prime(L)|^2= 2\Omega_I$. 
The
real part, $\Omega_R$, gives an extra phase factor to the parallel
component. 
The integral in the real part of $\delta\Omega$ can be done numerically.
We find,
\begin{equation}
\delta\Omega(L) \approx i{6\over 5}\pi^2C_N^2{l^2\over 4 \omega^2}
\left({gB_t\overline\omega^2_p\over 2n_e} \right)^2 Ll^{5/3}
+ {6\over 5}\pi C_N^2 {l^2\over 4 \omega^2} 
\left({gB_t\overline\omega^2_p\over 2n_e} \right)^2 4LlL_0^{2/3}
\label{DeltaOmega1}
\end{equation}

It is clear from Eq. \ref{DeltaOmega1} that the $\delta\Omega_R/\delta\Omega_I
\sim (L_0/l)^{2/3}$, where $\delta\Omega_R$ and $\delta\Omega_I$ are 
defined by $\delta\Omega = \delta\Omega_R + i \delta\Omega_I $.
Hence we find that $\delta\Omega_R$ is much larger than 
$\delta\Omega_I$ if the frequency is such that $l<< L$, $l<< L_0$
and $L_0<< L$.
We therefore find that for a wide range of frequencies which satisfy
these inequalities the phase
factor $\Omega_R >>\Omega_I$ irrespective of whether the plasma
density is uniform or has a fluctuating part. In both cases 
we find that the contribution to $\Omega_R \propto \omega$. 
For the Virgo supercluster, assuming that $C_N^2$ scales as 
$\overline n_e$ and $L_0$ scales as $L$ we find that  
$\delta\Omega_R\approx 9\times 10^5$ for optical frequencies
$\nu=10^7$ GHz and $g = 10^{-10}$ GeV$^{-1}$. Hence at optical
frequencies the real part of $\delta\Omega$ is comparable to
the imaginary part. However at lower frequencies the real part is
much larger.

We next consider the case when the source is emitting axions along with
photons. In this case $\phi^\prime(0) \ne 0$. The expression for 
$A_\parallel^\prime(L)$ can now be written as
\begin{eqnarray}
A_\parallel^\prime(L) &=& A_\parallel^\prime(0) - i{gB_t\over 2}\phi^\prime(0)
\int_0^L dz\exp\left(-i{\xi\over 2\omega}\right)\nonumber\\
&-& \left({gB_t\over 2}\right)^2 A_\parallel^\prime(0)
\int_0^L dz\exp\left(-i{\xi\over
2\omega}\right)\int_0^z dz^\prime
\exp\left(i{\xi\over 2\omega}\right)
\label{phine0}
\end{eqnarray}
In order to compute the polarization properties of the electromagnetic
wave after propagation through a large distance $L$ we need to compute 
the correlations $<A_\parallel^{\prime*}(L)A_\parallel^\prime(L)>$
and $<A_\perp(L)A_\parallel^{\prime*}(L)>$. We ignore all terms
which involve third and higher powers of the coupling $g$. Furthermore
we include only those terms which scale like $L$ for very large
values of $L>> l$. With these requirements it is clear that only three 
terms contribute to $<A_\parallel^{\prime*}(L)A_\parallel^\prime(L)>$. These
are $|A_\parallel^\prime(0)|^2$, 
\begin{equation}
\left<\left|i{gB_t\over 2}\phi^\prime(0)\int_0^L dz\exp\left(-i{\xi\over 2\omega}\right)\right|^2\right>\ , 
\label{second}
\end{equation}
and
\begin{equation}
\left<-A_\parallel^{\prime*}(0)\left({gB_t\over 2}\right)^2 A_\parallel^\prime(0)
\int_0^L dz\exp\left(-i{\xi\over
2\omega}\right)\int_0^z dz^\prime
\exp\left(i{\xi\over 2\omega}\right)+ c.c.  \right>\ .
\label{third}
\end{equation}
The second term (Eq. \ref{second}) given above is equal to the result
given in Eq. \ref{Prob} upto an overall factor of $|\phi^\prime(0)|^2$.
The third term, given in Eq. \ref{third}, is equal to the RHS of 
Eq. \ref{EqOmega} (plus its complex conjugate) up to an overall factor of
$iA_\parallel^{\prime*}(0)A_\parallel^\prime(0)$. The fluctuating
part of this expression is evaluated in Eq. \ref{DeltaOmega1}.
Similarly the only nontrivial
term that contributes to $<A_\perp(L)A_\parallel^{\prime*}(L)>$
involves the expectation value of the third term on the RHS of Eq. \ref{phine0} 
which is already evaluated in Eq. \ref{DeltaOmega1}. 
We point out that the expectation value of the second term on the RHS of
Eq. \ref{phine0} does not scale like $L$ in the limit of large $L$.
Hence the terms which involve interference of this term with a term
independent of fluctuations is neglected.
We, therefore, find that the only change introduced due to $\phi^\prime(0)\ne 0$
is an extra term in $<A_\parallel^{\prime*}(L)A_\parallel^\prime(L)>$
which arises due to decay of axions. This is the term given in 
Eq. \ref{second} and upto an overall factor of $|\phi^\prime(0)|^2$,
its expression is given in Eq. \ref{Prob}. 

We next compute the changes in polarization induced by the phase
$\Omega_R$ due to axion photon 
mixing. We neglect the imaginary part of $\Omega$ in the following
discussion. As shown above it is in general much smaller than the
real part. Furthermore its effect can be easily included by multiplying
the component of vector potential (or the electric field) 
parallel to $\vec B_t$ by $e^{-\Omega_I}$. 
If the incident electromagnetic 
wave is unpolarized then the phase $\Omega_R$ does not produce any
additional effect.
However if the incident beam is already polarized
then $\Omega_R$ can change also its polarization. Let's consider the simple
case of a 
monochromatic wave,
\begin{eqnarray}
E_x &=& a_1 \exp(i\alpha_1-i\omega t)\nonumber\\
E_y &=& a_2 \exp(i\alpha_2-i\omega t)
\end{eqnarray}
where $E_x$ and $E_y$ are the two transverse components of the 
electromagnetic wave.
By choosing the coordinate axis such that the y-axis is aligned along 
the direction of the transverse component of the 
background magnetic field, we find that 
after propagation through the medium the y component of the
electric field $E_y$ acquires an additional phase $\Omega_R$. Hence the final
values of the Stokes parameters are given by
\begin{eqnarray}
S_0 & =& a_1^2 + a_2^2\\
\label{s0}
S_1 & =& a_1^2 - a_2^2\\
\label{s1}
S_2 & =& 2a_1 a_2\cos(\Delta + \Omega_R)\\
\label{s2}
S_3 & =& 2a_1 a_2\sin(\Delta + \Omega_R)
\label{s3}
\end{eqnarray}  
where $\Delta = \alpha_2 - \alpha_1$.
The linear polarization angle $\psi$ is then given by
\begin{equation}
\tan 2\psi = {S_2\over S_1} = {2a_1a_2\cos(\Delta+\Omega_R)\over a_1^2-a_2^2}
\label{psi}
\end{equation}
The degree of polarization $P=(S_1^2+S_2^2+S_3^2)^{1/2}/S_0$ remains
equal to unity. Similarly the degree of circular polarization
defined as $P_C = |S_3|/S_0$ is given by
\begin{equation}
P_C=\left|2a_1 a_2\sin(\Delta + \Omega_R)\right|/(a_1^2 + a_2^2)
\label{pc}
\end{equation}
and the degree of linear polarization 
$P_L = (S_1^2+S_2^2)^{1/2}/S_0$ is given by
\begin{equation}
P_L= [(a_1^2 - a_2^2)^2 + 4a_1^2 a_2^2\cos^2(\Delta + \Omega_R)]^{1/2}/S_0\ .
\label{pl}
\end{equation}
We see that the linear polarization angle $\psi$, the degree of 
circular polarization $P_C$ and the degree of linear polarization $P_L$ all
depend in a precise manner on the frequency $\omega$ of the incident 
wave. If we assume that $\Omega_R$ is small then these depend linearly
on $\omega$. In general these are oscillatory functions of $\omega$. 
On the other hand the degree of polarization $P$ is independent of
frequency. Hence we can easily check this effect by analyzing spectral
data from distant AGNs at optical and higher frequencies. 

If the incident beam is quasi-monochromatic the above analysis 
goes through with minimal modification. The wave now need not
be completely polarized. We may write its coherency matrix as
\begin{equation}
{\bf J} = \left(\matrix{J_{11} & J_{12}\cr J_{21} & J_{22}}\right)
\end{equation}
where $J_{ij} = <E^*_i(t) E_j(t)>$ and $J_{21} = J_{12}^*$ \cite{Mandel}. After 
propagating through the magnetic field which is assumed to be pointing
in the $\hat y$ direction we find that the y-component $E_2$ of the
electric field acquires an additional phase $\Omega_R$.
Hence the coherency
matrix changes to 
\begin{equation}
{\bf J} = \left(\matrix{J_{11} & e^{i\Omega_R}J_{12}\cr e^{-i\Omega_R}J_{21} 
& J_{22}}\right)
\end{equation}
The Stokes parameters can be written as $S_0= J_{11}+J_{22}$,
$S_1= J_{11}-J_{22}$, $S_2= e^{i\Omega_R}J_{12} + c.c.$ and
$S_3 = i(e^{-i\Omega_R}J_{21}- c.c.)$. Parametrizing $J_{12}=je^{i\delta}$ 
we find that $S_2 = 2j\cos(\delta+\Omega_R)$ and $S_3 = 2j\sin(\delta+
\Omega_R)$.
We again find that the degree of polarization is independent of
$\Omega$ but the orientation of linear polarization as well as the
degree of linear and circular polarization ($P_L$ and $P_C$) become
oscillatory functions of $\omega$ due to their dependence on $\Omega_R$.

\section{Astrophysical Applications}
We next examine if the astrophysical effects discussed in the introduction
can be explained in terms of axion photon mixing. As we have seen the
effect is too small for the interstellar magnetic fields. However the
effect can be large and produce observable consequences for the case 
of Virgo supercluster magnetic fields, especially at optical frequencies.
We first consider the polarization alignment effect claimed by
Hutsemekers \cite{Hutsemekers}. This effect can be explained by
the depletion of the photons polarized parallel to $\vec B_t$.
We point out that the real part of $\Omega$ by itself cannot explain this
effect. 
As explained 
in the introduction, a very striking alignment was found in the region
A1 in ref. \cite{HL}, 
which is bounded in Right
Ascension by
$11^{\rm h}15^{\rm m}
\le {\rm RA} \le 14^{\rm h}29^{\rm m}$ and in redshift by $1.0\le z\le 2.3$.
In Ref. \cite{HL} it is also shown that the quasar polarization in 
the region A1 show alignment with the supergalactic plane.
By invoking
coherent cluster magnetic field it may be possible to explain the existence of
large scale correlation. 
The region A1 turns out to be in the direction
of the Virgo supercluster and this appears to be a promising explanation. 
As pointed out in Ref. \cite{HL} the correlation with the supergalactic
plane may be evidence in favor of a propagation effect such as
axion photon mixing. An alternate possibility, also mentioned in 
Ref. \cite{HL},
is extinction due to dust. However this explanation suffers from two 
drawbacks. The first problem is that the 
effect is redshift dependent, i.e. the
regions which show alignment are bounded in redshift as well as in
angular coordinates. Secondly one finds that in general the distribution of 
the degree of polarization for the radio quiet (RQ) along with 
optically selected non-broad absorption line quasars (O) differs from the
distribution of the broad absorption line (BAL) quasars. 
Precisely the same difference between
these different types of quasars is also seen in the A1 region \cite{HL}. 
If the supercluster magnetic field causes the alignment in polarization
due to decay of $A_\parallel$
then this difference would not be preserved in the A1 region. 

It, therefore, seems that a propagation effect is not likely to
provide an explanation for this effect. This would, however, imply that
the quasar polarizations are intrinsically aligned with one another
over cosmologically large distances.  
Correlations over cosmological distances
violate the basic assumption of isotropy and homogeniety
of the Universe.
Here we propose an alternate mechanism which
explains these observations in terms of a propagation effect related
to axion photon mixing. 
We first notice that there are very few RQ+O quasars in the A1 region
which satisfy
the cut $p>0.6$ \% imposed on the data \cite{Hutsemekers}. Hence
these quasars do not contribute significantly to the alignment effect. 
We next assume that the quasars are emitting axions
along with the photons. 
During propagation the axions decay into photons 
in the presence of background magnetic field. The decay probability of
axions is also given by Eq. \ref{Prob}. The emitted photon is polarized
in the direction parallel to the transverse magnetic field. 
The axion photon coupling is assumed to be such that in the
optical frequencies the decay probability of axions, or that of photons, 
from distant quasars
is small, less than about 1 \%. The total photon flux generated
due to decay of axions is equal to the product of the initial axion
flux $\Phi$ and its decay probability $P_{\phi\rightarrow \gamma}$. 
We further assume that for the BAL quasars the initial axion flux 
$\Phi$ is large enough at optical frequencies such that it leads
to significant change in polarization during propagation. In
order to explain the Hutsemekers effect we require that the axion
flux is of the same order of magnitude as the photon
flux from BAL quasars, 
if the probability $P_{\phi\rightarrow \gamma}\approx 1$ \%. 
In the current paper we will not 
address the question of emission rate of axions from quasars
since the basic physics of the interior
of the quasars is not well understood. 
For the (RQ+O) quasars
we assume that the emitted axion flux is small enough such that
it does not affect the optical polarizations. The distribution of
the degree of polarization for the (RQ+O) quasars will, therefore, be
unaffected due to axion photon mixing. However the degree of polarization 
for the case of BAL quasars will be determined primarily by the propagation
effects. Hence this mechanism explains the observed difference between
the BAL and (RQ+O) quasars.

The redshift dependence of the Hutsemekers effect \cite{Hutsemekers,HL}
is difficult to explain even with the axion decay mechanism discussed
above. However we notice that the dominant contribution to the effect
comes from the A1 region which can be explained in terms of the Virgo 
supercluster magnetic fields. The strong alignment of the polarizations
with the supercluster plane, as shown in Fig. 3 of Ref. \cite{HL},
provides further support for this explanation. Some redshift dependence
may be obtained due to the evolution of the quasars with redshift. 
It is possible that the axion emission rate of quasars is redshift
dependent and hence only quasars within a certain redshift interval
give dominant contribution, with the remaining quasars dominantly
contributing noise. 
It is also possible that the dominant contribution
to the effect comes from a few clusters within
our astrophysical neighbourhood which can then also contribute to the
observed redshift dependence.
This can be studied by collecting a larger data 
sample and putting appropriate cuts on the redshift.

\begin{figure}
\psfig{file=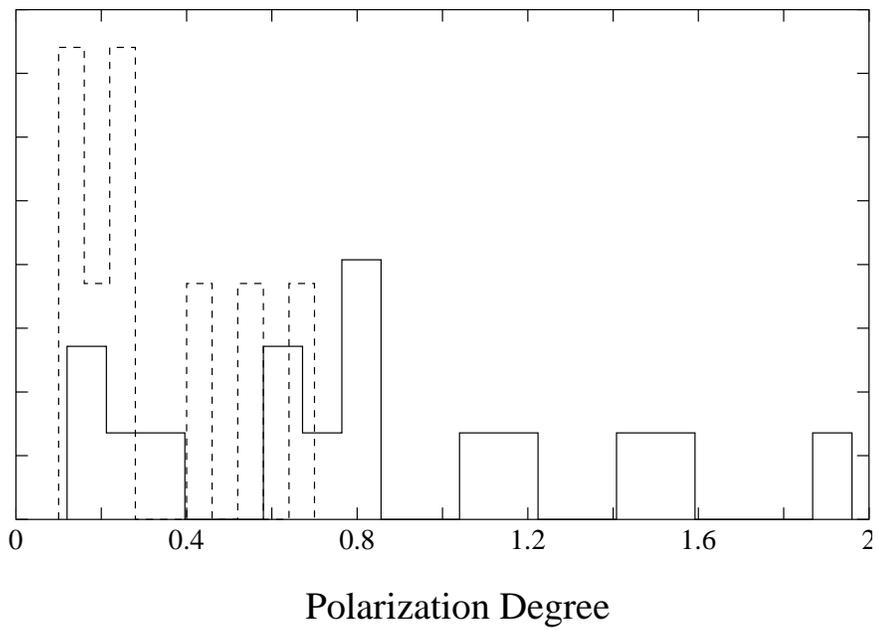}
\caption{The distribution of polarization degree $p$ (in \%) 
for the BAL quasars
(solid line) and the RQ+O quasars (dashed line) for a patch of $20^o$
radius centered around the Virgo supercluster.}
\label{distribution}
\end{figure}

It is interesting to note that the observed difference between
the BAL and RQ+O quasars in the direction of the Virgo supercluster
allows us to put a very stringent constraint on the pseudoscalar photon
coupling $g$ in the limit of $m_a<<\overline\omega_p$ provided we 
assume our extrapolated value of the correlation coefficient $C_N$. 
We may isolate the influence of the Virgo supercluster by taking only
the region in which a large Faraday Rotation Measure is observed. This
corresponds to a patch of $20^o$ radius around the center of
the Virgo supercluster (RA=$12^{\rm h}28^{\rm m}$, DEC=$12^o40^\prime$) 
\cite{Vallee}. 
We find that in this region there are
16 BAL quasars and 9 RQ+O quasars among the objects given in table 2
of Ref. \cite{HL}. The difference between these two types of quasars persists
in this region also as can be seen in Fig. \ref{distribution}.
If the photon to axion conversion probability is large, i.e. 
$P_{\gamma\rightarrow \phi} > 0.1$, 
then this difference
would be completely washed out due to propagation through the Virgo 
supercluster. This is because the degree of 
polarization $p$ of RQ+O quasars peaks at
very small values ($p\approx 0.2$ \%) and we find no quasars in this
category with $p>2$ \%.  
This observation implies that the photon decay probability $P_{\gamma
\rightarrow \phi}<1$, where the upper limit, unity, is chosen in order to
obtain a conservative upper bound on $g$. 
By using Eq. \ref{Prob} and the parameters
for the Virgo supercluster, given in the discussion following Eq. \ref{Prob},
we find that 
\begin{equation}
g < 10^{-13}\ {\rm GeV}^{-1} \left[{3.4\times 10^{-14}\ {\rm m}^{-20/3}
\over C_N^2}\right]\left[{1\ \mu {\rm G}\over B_t}\right]
\left[{\overline n_e\over 10^{-6}\ {\rm cm}^{-3}}\right]
\left[{10\ {\rm Mpc}\over L}\right]. 
\end{equation}
The reliability of this constraint depends on our assumed extrapolation 
of the correlation coefficient $C_N$ for the supercluster. 
In future it is clearly of interest to obtain this parameter by
direct observations which can lead to a more reliable constraint.

In the radio frequencies ($\sim 1$ GHz) the effect is clearly very small.
The dominant contribution is given by 
$\delta\Omega_R$ assuming Virgo supercluster parameters, given in the
discussion following Eq. \ref{DeltaOmega1}. 
We find that for radio frequencies $\nu = 1$ GHz, $\delta\Omega_R\approx 0.1$
for $g=10^{-10}$ GeV$^{-1}$ with $m_a<< \overline \omega_p$. 
The appropriate value of $g$ is, however, likely to
be much smaller than $g=10^{-10}$ GeV$^{-1}$. 
We next investigate whether the proposed emission
of axions due to quasars can explain this effect.
Given a large enough flux of
axions it is clearly possible to obtain an observable effect on the 
polarization at radio frequencies. The probability for decay of 
axions into photons $P_{\phi\rightarrow \gamma}$ in the presence of
background magnetic fields is also given
by the result given in Eq. \ref{Prob}.  
Using the CMBR constraint on the coupling $g$ we find that the 
$P_{\phi\rightarrow \gamma}\le 6.5\times 10^{-7}$
at radio frequencies $\nu \sim 1$ GHz. 
Hence in order that axions decay can lead to observable consequences
the axion flux has to be roughly $10^6$ times
larger than the photon flux $F_\gamma$ at radio frequencies.
The photon flux at radio frequencies is given by $C/\nu^q$ where $C$
is a constant and the spectral index $q$ varies roughly between 0.5 to
2 and is in general larger for larger frequencies. 
Assuming that the axion flux from quasars is indeed $10^6$ times
larger than the photon flux at radio frequencies and that its spectral
dependence
is given by $A/\nu^p$, 
where  $A$ and $p$ are constants, we find $p\ge q+5/3$ in order that 
photon spectrum is not in conflict with observations. Assuming that $q\approx
0.8$ at radio frequencies we find that $p\approx 2.5$. The axion luminosity of
the quasar is given approximately by $4\pi L^2A/[(p-1)\nu_c^{p-1}]$,
where $\nu_c$ is the lower cutoff on the frequency and $L$ is the total
distance of propagation.
We, therefore, find that the axion luminosity is larger than
 roughly $10^6$ 
times the radio photon luminosity. For most quasars this is one to two 
orders of magnitude larger than their total bolometric luminosity. 
This large luminosity required suggests that this mechanism is disfavored
and we do not pursue it further.

Another mechanism that might be relevant for explaining the 
anisotropy in radio waves involves the presence of background
axion field. As shown in Ref. \cite{HS92} the rotation in the linear
polarization angle due to a background magnetic field is equal
to $g\Delta\phi/2$ where $\Delta\phi$ is the total change in the 
background axion field along the trajectory of the electromagnetic
wave. If we assume that the axion field in our neighbourhood 
shows a dipole distribution, $\phi(\vec r) = \phi_0(r) \cos(\Theta)$
where $\Theta$ is the angle with respect to the dipole axis and
$\phi_0(0)$ is of order $1/g$, then the dipole anisotropy claimed
in Ref. \cite{Birch,JR} can be explained. This explanation requires
that the axion field at very large distances, which correspond to the positions
of the quasars, is relatively smooth. The random component of the
axion field should have a strength smaller than $1/g$. However it
is difficult to theoretically justify the existence of such a dipole
distribution of the axion field and hence this explanation is also
not very compelling. 
 
Finally we examine the recent claim \cite{Csaki} 
that the observed supernova dimming \cite{Perlmutter,Riess}
at large redshifts can be explained in terms of axion photon mixing. 
It has been pointed out that axion photon mixing is unlikely
to provide an explanation because inclusion of the intergalactic
plasma density leads to a considerable reduction of the effect
\cite{Harari}. Furthermore it also gives rise to a spectral dependence
which is incompatible with observations \cite{Harari}.
The intergalactic parameters used in Ref. \cite{Harari} are 
$B=10^{-9}$ Gauss and $\overline n_e=10^{-7}$ cm$^{-3}$. However in
Ref. \cite{Csaki2001} the authors argue that the intergalactic
plasma density is
likely to be much smaller than this. By using $\overline n_e=10^{-8}$
cm$^{-3}$ it is found that the axion photon mixing can provide
an explanation for the dimming effect \cite{Csaki2001,Mortsell}.
Here we examine the magnitude of the effect due to the fluctuating
part of the plasma density as given in Eq. \ref{Prob}. Here again
if we assume that $L_c<< l$ then the $l$, and hence the spectral, 
dependence goes away. The result in this case can be obtained 
by calculating $|\phi^\prime(L_c)|^2$ which is the probability
$P_{\gamma\rightarrow \phi}$ over a single coherence length
and then adding all the contributions over the entire propagation distance.
The decay probability over a single coherence length is given by
\cite{CG94}
\begin{equation}
P_{\gamma\rightarrow \phi} = \left[{gB_t\overline\omega_p^2\over 2
\overline n_e (\overline\omega_p^2-m_a^2)}\right]^2 <I>
\end{equation}
where 
\begin{equation} 
I = \left|\int_0^{L_c} dz \delta n_e \exp\left(i {\overline\omega_p^2-m_a^2
\over 2 \omega} z\right)  \right|^2\ .
\end{equation}
We then find, using $l>>L_c$, that 
\begin{equation}
<I> = 4\pi \int_{-\infty}^\infty dk_z\int_0^\infty dk_T^2
{\sin^2(k_zL_c/2)\over k_z^2} P_{3N} 
\end{equation}
Here $P_{3N} = C_N^2/(k_z^2 + k_T^2 + 1/L_c^2)^{11/6}$ where the outer
length $L_0$ has been set equal to the coherence length. We perform  
the $k_T$ integral to find
\begin{equation}
<I> = {24\over 5}\pi C_N^2 {L_c^{8/3}\over 2^{5/3}} 
\int_0^\infty dx
{\sin^2 x\over x^2} {1\over (x^2+1/4)^{5/6}} 
\end{equation}
The remaining 
integral can now be done numerically and is found to be approximately
equal to 2. It is clear that the 
effect is independent of $\nu$. Adding the contributions from the 
entire propagation distance we need to replace $L_c^{8/3}\rightarrow
LL_c^{5/3}$. Here we have assumed that the total effect is small and
the initial axion flux is negligible. 
We have also ignored the fact that magnetic fields in different domains
are aligned in different directions. A more reliable summation over
different domains in obtained in Ref. \cite{Roy}. Here we are only
interested in an order of magnitude estimate and hence these additional
refinements are unnecessary. 
In order to estimate the magnitude of the effect 
we take $L = 1$ Gpc,  $L_c=0.1$ Mpc, $g= 10^{-11}$ GeV$^{-1}$, 
$n_e = 10^{-7}$ cm$^{-3}$ and again
assume that $C_N^2$ scales as $n_e^{20/9}$. We point out that the parameters
$L_c$, $g$ and $C_N$ are subject to considerable uncertainty. We have
chosen a range which is allowed by current observations and for 
which the effect is significant and independent of $\nu$. 
With these choice of
parameters we find that $P_{\gamma\rightarrow \phi}\approx 1.5$,
i.e. the conversion probability is of order unity.
It is clear that the range of allowed parameter space is even larger
if we use $n_e = 10^{-8}$ cm$^{-3}$. 
Hence we cannot exclude the possibility that axion photon mixing 
may be responsible for the dimming of distance supernovae.

\section{Conclusions} 
In conclusion we have analyzed the mixing of axions with photons. We
have determined how the dispersion relations of the electromagnetic
waves are modified due to the presence of axions. We found
that for a wide range of parameters this provides the dominant
contribution to the changes in polarization of electromagnetic waves
due to mixing with axions. We have also determined whether the existence
of axions can explain the large scale anisotropies claimed in References
\cite{Birch,JR,Hutsemekers}. We find that the Hutsemekers \cite{Hutsemekers} 
effect
may be explained by the supercluster magnetic fields if we assume that
quasars emit axions such that their flux is of the order of or larger
than the photon
flux at optical frequencies. The Birch  
effect \cite{Birch,JR} may also be explain by assuming axion emission
from radio galaxies and quasars but requires a very large flux at radio
frequencies. Its explanation in terms of axions is therefore disfavored. 
Finally we have estimated the contribution of the fluctuations in the
plasma density to the 
dimming of distant supernovae due to
axion photon mixing. We find that this effect can be rather large and for
a considerable range of currently allowed parameter space can provide
an explanation for the supernova dimming.

\bigskip
\noindent
{\large\bf Acknowledgements:} We thank John Ralston, Ajit Srivastava
and Mahendra Verma
for very useful comments. This work is supported in part by a grant from 
Department of Science and Technology, India.

\end{document}